\documentstyle[preprint,aps,psfig]{revtex} %eqsecnum
\newtheorem{definition}{Definition}
\newtheorem{proposition}{Proposition}
\newtheorem{properties}{\sc Property}
\newtheorem{corollary}{Corollary}
\newtheorem{exemple}{\sc Exemple}

\def\be{\begin{equation}}
\def\ee{\end{equation}}
\def\bea{\begin{eqnarray}}
\def\eea{\end{eqnarray}}

\def\bdf{\begin{definition}}
\def\edf{\end{definition}}
\def\bpr{\begin{properties}}
\def\epr{\end{properties}}
\def\bpt{\begin{proposition}}
\def\ept{\end{proposition}}
\def\bcll{\begin{corollary}}
\def\ecll{\end{corollary}}
\def\bex{\begin{exemple}}
\def\eex{\end{exemple}}

\begin{document}

%Pour changer les labels des footnotes des numeros a symbols:
\renewcommand{\thefootnote}{\fnsymbol{footnote}} 
\def\scf{\setcounter{footnote}}
% example: \scf{1}\footnote{}

%\setcounter{page}{113}
%\thesaurus{06(02.04.1; 02.18.8; 08.14.1; 08.16.6; 08.18.1; 08.05.3)}
\hyphenation{coor-don-nees Ex-pli-cite-ment res-pec-tive-ment Le-gen-dre 
pro-pa-ga-tion va-ria-bles}
\title{Galactic periodicity and the oscillating $\mbox{\boldmath{$G$}}$ model
%Oscillating gravitational constant and periodic distribution of 
%galaxies
}

\author{ Marcelo Salgado\footnote{e-mail: 
marcelo@roxanne.nuclecu.unam.mx}, 
Daniel Sudarsky, and Hernando Quevedo \\ 
\small {\it Departamento de Gravitaci\'on y 
Teor\'\i a de Campos, Instituto de Ciencias Nucleares} \\ \small
{\it Universidad Nacional Aut\'onoma de M\'exico} \\ \small {\it 
Apdo. Postal 70-543 M\'exico 04510 D.F, M\'exico} }
\maketitle
\centerline{ (Received 21 December, 1995)}
We consider the model involving the oscillation of the
effective gravitational constant that has been put forward in an attempt to 
reconcile the observed periodicity in the galaxy number distribution with the 
standard cosmological models. This model involves a highly nonlinear dynamics 
which we analyze numerically. We carry out a 
detailed study of the bound that 
nucleosynthesis imposes on this model. The analysis shows that for any 
assumed value for $\Omega$ (the total energy density) one can fix the value of 
$\Omega_{\rm bar}$ (the baryonic energy density) 
in such a way as to accommodate the observational constraints coming 
from the $^4{\rm He}$ primordial abundance. In particular, if we impose the 
inflationary value $\Omega=1$ the resulting baryonic energy density turns out 
to be $\Omega_{\rm bar}\sim 0.021$. This result lies in the very narrow range 
$0.016 \leq \Omega_{\rm bar} \leq 0.026$ allowed by the observed values of the 
primordial abundances of the other light elements. The remaining fraction of 
$\Omega$ corresponds to dark matter represented by a scalar field. 
 \vskip 1.5cm
\noindent PACS number(s): 98.65.Dx, 04.50.+h, 95.35.+d, 98.62.Py

%\noindent {\bf Key words:}{gravitational constant, scalar field, 
%non-minimal coupling}
%Submited to Computer Physics Communications in February 1993.%
\newpage
%running head title: {\bf oscillating $G$ models}

%fax: 905-6.16.22.33

%telephone number: 905-6.22.46.72

%e-mail: marcelo@roxanne.nuclecu.unam.mx

\newpage

\section{Introduction}
\bigskip
The recent  observations, in deep pencil beam surveys \cite{BEKS}, 
showing that the galaxy number distribution exhibits a remarkable 
periodicity of 128$h^{-1}\;$Mpc comes as a shocking development, since, if taken at face value it would imply that we live in the middle of a pattern
consisting of concentric two-spheres that mark the maxima of the 
galaxy number density.
This, of course, would  be catastrophic for our cosmological conceptions.
While it is true that such periodicity has been observed only in the few 
directions that have been explored so far, it would be a striking 
coincidence if it turns out that it is absent in other directions and we 
just happen to have chosen to explore the only directions in which that 
phenomenon occurs. Therefore 
it seems reasonable to assume that the periodicity is also present 
in the deep pencil beam surveys in
other directions, thus taking us to the concentric spheres scenario.
The seriousness of the situation is such that this type of scenario has indeed been
put forward in a model where the formation of these concentric shells is 
a result of a ``spontaneous breakdown of the cosmological principle'' 
via a mechanism that results
in the appearance of patches filled with the concentric spheres pattern, with these
patches covering the Universe \cite{Bun}. Again it would still be difficult to explain how we did come out living in the center of such a patch (more precisely inside 
the innermost sphere of one such patch).

The only known way out of this type of scenario is to assume that there is 
only an
apparent spatial periodicity that is the result of a true temporal periodicity which
 shows up in our observations of distant points in the Universe and that is mistakenly
interpreted as a  spatial periodicity \cite{Hill}. The models that have been put forward in order
to achieve this temporal periodicity involve the oscillation of an effective coupling
constant due to a contribution to it coming from the spectation value of some scalar field,
that actually oscillates coherently in cosmic time in the bottom of its effective potential.

The specific models that have been proposed involve the oscillation of the
 effective 
electric charge, electron mass, galaxy luminosity or gravitational constant
\cite{Hill,Morik,CritStein,SisVuc}. From these the first two have been shown to conflict 
with bounds arising from the test 
of the {\it equivalence principle}\cite{Sud}. As for the third scenario, 
it  would
seem to involve a large number of hypotheses since the galactic luminosity is fixed by
the number and type of stars present in the galaxy and their respective luminosities, and
the latter  are themselves functions of the standard  physics coupling constants that control nuclear reaction rates (on the variation of which there are 
severe limits), and of the transport mechanisms: convection, radiation, etc., which are also
determined by the standard physics coupling constants. Thus it seems that
the only way to produce such a model requires the introduction of ``exotic'' 
particles (like axions or massive neutrinos \cite{Hill}) 
that would act as a new transport
mechanism and besides that, the hypothesis of a second mechanism that 
would produce the oscillation, an assumption which presumably involves the 
coupling of these hypothetical particles  to the hypothetical cosmological
scalar field.

\medskip

In light of the complicated nature of the alternative scenario,
it seems worthwhile to carry out a careful analysis of the viability of the
oscillating gravitational constant model, despite the 
difficulties that seem to appear when confronting the predictions 
of the model with other experimental data. We will address these 
difficulties below.

\medskip

The oscillating $G$ model is
based on a  cosmological, massive scalar free field, that is
nonminimally coupled to curvature and whose oscillations in cosmic time 
result in the oscillation of the effective gravitational constant. 
The model is governed by a system of nonlinear, ordinary,
differential equations that we shall integrate numerically without 
analytical simplifications. 
In previous related works\cite{CritStein,Hill,Morik}, 
only approximate solutions have been 
explored as in the work 
of Morikawa\cite{Morik}. In this work, the dynamical equation for the scalar field 
is linearized into a Bessel equation. From this, 
the Hubble parameter for a flat matterless Universe is evaluated. 
The approximations result in a precision of 2\% in the Hamiltonian 
constraint. While such procedure is an extremely simple way to study 
the implementation of the model it is however not totally satisfying.
 In addition to the 
analytical simplifications Morikawa made, we emphasize that his ``matterless''
 (no baryonic or radiation energy density) Universe assumption is not 
useful for the study of the constraints we discuss below, 
notably the one imposed by 
nucleosynthesis. As we shall analyze in Sec. IV, it is actually 
the presence of matter that makes it possible to satisfy such a constraint. 
 Furthermore, 
it is obvious that Morikawa's approach is no longer valid in early times where 
the linear approximations break down. Hence, this method cannot be 
used to describe the behavior of cosmological variables during the 
nonoscillating phase.

The data arising from the Viking radar echo experiments in addition to those 
related with the limits on the Brans-Dicke parameter \cite{CritStein}
impose certain bounds on the value of $\dot G/G$. There is an 
apparent conflict between these bounds and the value they should have 
for reproducing the galaxy-amplitude counting. However,
this problem can be overcome by imposing the unnatural extra requirement that our 
Galaxy is at a ``fortunate phase" \cite{CritStein} where the scalar field is 
swinging very close to zero at our particular place and time in the 
Universe. 
There are additional bounds on the oscillating $G$ model arising
from Big Bang nucleosynthesis,  but we will argue that
these cannot be adequately addressed in the fashion described by Accetta 
{\it et al.} \cite{Acetal}.
This constraint is obtained by considering the Helium abundance limits which
depend on the neutron-proton ratio at the temperature of freeze-out.
This temperature is itself obtained from the condition that the Hubble
 parameter equals the nucleon to proton weak reaction rate. 

Crittenden and Steinhardt \cite{CritStein}, based on a previous analysis by 
Accetta and collaborators that resulted in the bound 
$\Delta G /G < 0.4 $ \cite{Acetal}, argue that 
the nucleosynthesis constraint 
is so stringent that it practically rules
 out the oscillating $G$ model unless 
we assume a ``fine-tuning'' within the oscillations of the scalar field. 

The
problem with employing the analysis of Accetta {\it et al.} 
\cite{Acetal} directly to this model is that they study 
the change ocurring in the Hubble parameter as a function of the 
temperature when one changes the  value of $G$, but fails to take into
account the fact that the model implies an equation for the Hubble
parameter that contains terms other than the simple ones used 
by the authors. These terms are associated 
with the contributions from the scalar field to the 
effective energy density. 

In this paper we carry out an analysis of the nucleosynthesis bound 
taking into account these extra terms and demonstrate that indeed such a 
``fine-tuning'' is needed and moreover that it is possible. However, we will 
argue that this ``fine-tuning'' is not of the kind that should result in the 
dismissal of the model, but rather a natural adjustement of the initial 
conditions that will lead to the observational data extracted from our 
Universe today. In other words, scientific models that require a very precise
choice of the numerical value of the initial conditions in order to reproduce
a given qualitative behavior of the observational data, are models that would
be consider unnatural and the choice of the specific initial data is 
justifiably described as ``fine-tuning.'' However, models that require a very 
precise choice of the numerical value of the initial conditions
in order to reproduce a specific numerical observational data cannot be 
considered as unnatural, especially if for every conceivable value of the
observational data (at least in some range) there is a corresponding value 
of the initial data. In this type of models, the particular ``preferred'' value
at the initial data is just the result of a 1 to 1 correspondence between
initial conditions and final outcome. We will argue that the present model
could be of the latter type. 

%not directly related to the variation of $G$ since this era 
%but rather 

%In Sec. IV we put bound in the framework of the 
%Nevertheless, as we discuss in Sec. IV, the analysis they made 
%to arrive into that conclusion is not valid in this context. 
%The point is that the nucleosynthesis limit obtained 
%by Accetta et al. \cite{Acetal} and used by these authors 
%has not been appropriately put in the framework of an oscillating $G$ model  
%mediated by a scalar field and thus, such a bound cannot be 
%extrapolated to such scenarios as simply as Crittennden and Steinhardt  
%pretend \cite{CritStein}. Instead, it is a detailed analysis on the 
%value of the Hubble parameter at nucleosynthesis computed from the model  
%itself that can judge if this is consistent or not 
%with the nucleosynthesis limit. 

The organization of the paper is as follows: In Sec. II the model is 
described and the basic equations of evolution for the fields and matter are derived,
in Sec. III the numerical implementation is discussed together with the
error estimation analysis, in Sec. IV the results of the numerical 
integrations are analyzed %and compared with previous related works, 
and finally in Sec. V  we give a brief discussion of the main
features exhibited by the model, their physical significance and the 
overall viability of the model.
\bigskip
  
\section{Formulation of the model}

We will consider a model in which the effective 
gravitational constant becomes dependent on cosmic time due to 
a contribution to it coming from the spectation value of a 
scalar field. This can be achieved by considering  
a scalar field $\phi$ nonminimally coupled to gravity. One of the
simplest models of this kind is obtained by taking a Lagrangian as follows
\bea
{\cal L} = &&\left({ 1\over 16\pi G_0} + \xi \phi^2\right)
\sqrt{-g} R - \sqrt{-g} \left[ {1\over 2}(\nabla \phi)^2
+ V(\phi) \right] \nonumber \\
&+& {\cal L}_{\rm mat}\ .
\label{lag}
\eea
Here $G_0$ is the Newton's gravitational constant; $\xi$ stands for the 
nonminimally coupling constant; and $V(\phi)$ is a scalar potential to 
be specified later (see Sec. III). In this model we are also including 
an schematic matter Lagrangian ${\cal L}_{\rm mat}$. 
Equation (\ref{lag}) shows that
the introduction of the coupling term is equivalent to considering
an effective gravitational constant which explicitly depends
on the scalar field: 
\be
G_{\rm eff} ={G_0\over   1+ 16\pi G_0  \xi \phi^2}\ .
\label{geff}
\ee
The gravitational field equations following from the Lagrangian 
(\ref{lag}) can be written as
\be
R^{\mu\nu} - {1\over 2} g^{\mu\nu}R = 8\pi T^{\mu\nu}_{\rm eff}
\ee
where 
\bea\label{Teff}
T^{\mu\nu}_{\rm eff} &=& 
G_{\rm eff}\left(4\xi T^{\mu\nu}_\xi + T^{\mu\nu}_{\rm sf} +  
T^{\mu\nu}_{\rm mat}\right) \ ,\\
\label{feq}
T^{\mu\nu}_\xi &=& \nabla^\mu(\phi\nabla^\nu\phi) - g^{\mu\nu}
\nabla_\lambda (\phi \nabla^\lambda\phi) \ , 
\label{txi} \\
T^{\mu\nu}_{\rm sf} &=& \nabla^\mu\phi\nabla^\nu\phi
- g^{\mu\nu}\left[{1\over 2}  (\nabla \phi)^2 + V(\phi)\right] \ .
\label{tsf}
\eea

\noindent The energy-momentum tensor of ``matter'' $T^{\mu\nu}_{\rm mat}$ 
will be composed of a combination of two noninteracting perfect fluids, one 
corresponding to pure baryonic matter ($i=1$) 
and the other one representing a pure 
radiation field ($i=2$):
\be
T^{\mu\nu}_{\rm mat}= T^{\mu\nu}_{\rm bar} + T^{\mu\nu}_{\gamma} 
= \sum_{i=1,2}\left[ (p_i +e_i ) U^\mu U^\nu + p_i g^{\mu\nu}\right] \ ,
\label{tpf}
\ee
which posseses the symmetries of the spacetime. The scalar field 
will also be assumed to posses these symmetries. 

Finally, the equation of motion for the scalar field becomes
\be
\Box \phi + 2\xi \phi R = {\partial V(\phi)\over \partial \phi} \ .
\label{seq}
\ee

We will focus on the Friedmann-Robertson-Walker (FRW) 
spacetimes which describe isotropic and homogeneous 
cosmological models
\be
ds^2 = - dt^2 + a^2(t)\left[ {dr^2\over 1 - k r^2} + r^2d\theta^2
+ r^2\sin^2\theta d\varphi^2\right] \ ,  
\label{frw}
\ee
where $a(t)$ is the scale factor and $k = 1, 0, -1$. 

Our purpose is to study the behavior of the solutions of the 
gravitational, matter, and scalar field equations for the FRW line
element (\ref{frw}). Since these equations are highly nonlinear, 
it is a difficult task to find analytic 
solutions; therefore, we will approach the problem via a numerical analysis.

One of the equations we find is the Hamiltonian constraint 
\be 
\frac{{\dot a}^2}{a^2} + \frac{k}{a^2} = \frac{8}{3} \pi G_0 \,E \ . 
\label{ham}
\ee
The dynamical equation for the single gravitational degree of freedom is
\be 
\frac{\ddot a}{a} + 2 \frac{{\dot a}^2}{a^2}
+ 2 \frac{k}{a^2} = 4 \pi G_0\left(E - \frac{1}{3} S\right) 
\ ,  
\label{dyn}
\ee
where 
\be 
E = \frac{G_{\rm eff}}{G_0}\left[e + \frac{1}{2} {\dot \phi}^2 + V(\phi) 
- 12 \xi \phi \dot\phi \frac{\dot a}{a}\right] \ , 
\label{ene}
\ee
is the total effective energy density and 
\be
S = \frac{3G_{\rm eff}}{G_0}\left[ p + \frac{1}{2} {\dot \phi}^2 -  V(\phi)
    + 4 \xi \left( {\dot \phi}^2 - \phi \dot\phi \frac{\dot a}{a}
    - \phi \Box \phi \right) \right] \ .
\label{trace}
\ee
These source terms contain 
contributions from the three parts of the total energy-momentum
tensor given in eqs.(\ref{txi})--(\ref{tpf}).

Finally, the equation for the scalar field (\ref{seq}) can be 
written explicitly as
\be
\ddot\phi + 3 \dot \phi \frac{\dot a}{a} + 
\frac{\partial V(\phi)}{\partial \phi}
 =  16\pi G_0 \xi \phi ( E-S) \ ,
\label{seq1}
\ee
where we have replaced the scalar curvature in terms of the 
energy-momentum tensor quantities $E$ and $S$.

It is worth mentioning that the contribution of the scalar field 
to the expression $T^{\mu\nu}_{{\rm eff};\nu}= 0$ [see Eq. (\ref{Teff})] 
vanishes identically and thus the energy-momentum of the ordinary matter satisfies 
$T^{\mu\nu}_{{\rm mat};\nu}=0$. 
We will moreover assume that the 
two perfect fluid components (baryons and photons) do not interact 
among themselves, thus each of their corresponding energy-momentum tensors 
is separetely conserved leading to 
\be
\dot{e}_i + 3 (e_i + p_i ) \frac{\dot a}{a} = 0 \ .  
\label{claw}
\ee
 
\noindent Equation (\ref{claw}) integrates immediately with respect 
the scale factor like in the standard cosmology case. We find then
\bea
\label{inte}
 e &=& e_{\rm bar} + e_\gamma= c_1\left(\frac{a_0}{a}\right)^3 + 
c_2\left(\frac{a_0}{a}\right)^4 \,\,\,\,,\\
\label{intp}
p &=& p_{\rm bar} + p_\gamma= \frac{c_2}{3} 
\left(\frac{a_0}{a}\right)^4  \,\,\,\,
{\rm with} \,\,\,\, a_0:= a(t= t_0) \ .
\eea
Here we have assumed an equation of state
$p_\gamma=e_\gamma/3$ for the radiation part, 
whereas $p_{\rm bar}=0$ for the corresponding 
baryonic component. The first term of (\ref{inte}) represents 
thus the pure baryon energy density, while the second one the radiation 
contribution alone. The constants $c_1$ and $c_2$ are fixed by the 
``initial'' conditions (i.e., the conditions today). In particular (at $t=t_0$) we will be 
assuming a baryonic energy density of one tenth of the critical value 
(at least for the first numerical experiment, but latter we will use $c_1$
as an adjustment parameter) 
 and the 
radiation energy density corresponding to the 
$2.73$ K cosmic background radiation (CBR) (see also Sec. III). 
Hereafter we shall refer as to {\it matter} the combination of both fluids.

%For the numerical integration of the field equations 
With the aim of reducing the field equations into an initial value problem 
consisting of a system of first-order differential equations, we shall 
rearrange these conveniently. Here we present the final form of equations 
with source terms containing no second-order derivatives 
and introduce better-suited variables:
\be 
P_\alpha = - \dot \alpha (t)  = - \frac{\dot a}{a} \ ,
\qquad  P_\phi = \dot\phi \ ,
\label{not}
\ee
where
\be
\alpha (t) :=  \ln\!\left[ \frac{a(t)}{a_0}\right] \,\,\,\,.
\label{alpha}
\ee 
The dynamic equation (\ref{dyn}) then takes the form 
\be
{\dot P_\alpha} - {P_\alpha}^2 = \frac{4}{3}\pi\,G_0(E+S) 
\ ,
\label{dyn1}
\ee
where we have used the Hamiltonian constraint (\ref{ham}) in order
to eliminate from eq.(\ref{dyn}) the term proportional to $k$. 
Notice that $P_\alpha\equiv -H(t)$.

Introducing eqs.(\ref{not}) into the scalar field equation we
obtain
\be
\dot P_\phi - 3 P_\alpha P_\phi + \frac{\partial V(\phi)}{\partial \phi} = 
 16\pi\,G_0 \xi  \phi (E - S) \ .
\label{seq2}
\ee
The source terms then take the following form:
\be 
E = \frac{G_{\rm eff}}{G_0}\left[ e + \frac{1}{2}{P_\phi}^2 + V(\phi) 
+12\xi \phi P_\phi P_\alpha \right] \ ,
\label{ene1}
\ee
and 
\bea
S &=& \frac{3}{1+ 192\pi G_{\rm eff} \xi^2 \phi^2}\frac{G_{\rm eff}}{G_0}
\left[p + \frac{1}{2} {P_ \phi}^2 - V(\phi) 
 \right. \nonumber \\ 
 &+& \left. 4\xi \left(\phi P_\phi P_\alpha + {P_\phi}^2 
-\phi\,\frac{\partial V(\phi)}{\partial \phi}
\right) +64\pi G_0 \xi^2 \phi^2 \,E \right]\ .
\label{trace1}
\eea
In obtaining the source (\ref{trace1}) from (\ref{trace}) 
we have used the scalar field 
equation (\ref{seq2}) in order to eliminate the term with 
$\Box\phi$. 

%It's to be noticed that equation (\ref{seq2}) with source terms 
%(\ref{ene1}), (\ref{trace1}) and (\ref{geff}) can be recognized as the 
%equation (2.4b) of Steinhardt \& Will \cite{SteinWill} 
%with the choice $f= (1+ 16\pi G_0 \xi\phi^2)/16\pi G_0 $.??

Further analysis of this model requires the numerical integration of the 
field equations under appropriate initial conditions. 
This will be performed in the following sections. In the Appendix 
we provide the dimensionless form of the above equations.
 
\bigskip
\section{Initial conditions and fixing of parameters} 
\bigskip

The choice of our variables fixes in advance the values of 
$\alpha$ and $P_\alpha$ at present time. Then, the initial conditions 
are 
\bea
\alpha|_{t=t_0} &\equiv& 0 \,\,\,\,\,,\\
P_\alpha|_{t=t_0} &\equiv & -1 \,\,\,\,\,.
\eea
It is worth emphasizing that by {\it initial conditions} we will 
mean throughout the paper the 
value of the field variables at present time and not their corresponding 
value near the Big Bang.

As stated before, the initial condition for $e$ 
corresponds to the value of the 
(baryon plus radiation) energy densities today and then:
\be
e_0:= e|_{t=t_0} = e^{\rm bar}_0 + e^{\gamma}_0 = c_1 + c_2 \ .
\ee
We fix $c_2$ by the value $T= 2.73$ K from the CBR and choose 
for the moment a value of $c_1$ corresponding to $\Omega_{\rm bar}= 0.1$. 
Thus, the values of the constants appearing in Eqs. (\ref{inte}) and 
(\ref{intp}) are
\bea
c_1 &=& 0.1 e_c \ , \\
c_2 &\sim& 4.2\times 10^{-5} e_c \,\,\,\,
{\rm with} \,\,\,\, e_c := \frac{3c^2 H_0^2}{8\pi G_0}\,\,\,\,.
\eea
Here $e_c$ stands for the critial energy density in terms of the 
current value of the expansion rate $H_0$.

\noindent This choice will leave $\Omega$ (see below) as a free parameter which 
is adjusted in order to determine the initial conditions 
on $\phi$ which result in a model 
respecting the observed $^4 {\rm He}$ abundances. 
As we mentioned before, we will be able to follow (Sec. IV) a 
different strategy that consists in imposing  $\Omega=1$ and using  $c_1$ 
(i.e., $\Omega_{\rm bar}$) as an adjustement parameter. 

The Viking data experiments constraint  
$\dot G/(GH)|_{\rm today} $ to be less than $0.3h^{-1}$ \cite{CritStein}. 
This imposes a 
bound on $\dot \phi$. The most conservative and the one we choose is 
\be
\dot \phi|_{t=t_0}= 0 \,\,\,\,\,\,.
\ee
However is straightforward to explore a less conservative initial condition. 

The values of the initial scalar-field's amplitude 
and coupling constant are obtained from the observational data. 
The observed redshift-galaxy-count amplitude 
(see \cite{CritStein,Hill}) and the 
Hamiltonian constraint (\ref{ham}) at $t_0$ provide two 
algebraic equations which determine the values of $\phi_0$ and $\xi$ 
once we choose a value for $\Omega$ and $\Omega_{\rm bar}$. When assuming a 
harmonic scalar potential 
$V(\phi)= m^2\phi^2$, these equations are
\bea
{\cal A}_0 &=& - \frac{G'_{\rm eff} \omega \phi_0}{2H_0 G_{\rm eff} }\,\,\,\,\,,\\
\phi_0^2 &=& \frac{e_0 -\Omega}{16\pi\xi \Omega - m^2} \,\,\,\,,
\eea
where $e_0$ is the initial matter energy density in units of 
$e_c$ ; $\Omega:= E_0/e_c$ and $\omega^2 \propto m^2$ is the 
oscillation frequency in units of $H^{-1}_0$.

Solving for $\phi_0$ and $\xi$ we obtain:
\bea
\xi &=& \frac{{\cal A}_0 m^2}{16\pi \left[{\cal A}_0 e_0 + 
\omega \left(\Omega - e_0\right) \right]}
\,\,\,\,,\\
\phi^2_0 &=& \frac{{\cal A}_0}{16\pi\xi\left[\omega - {\cal A}_0\right]} \,\,\,\,.
 \eea

The model will explain the observed galaxy distribution if 
${\cal A}_0\geq 0.5$ \cite{Hill}. 
The remaining parameter to be adjusted is 
$\omega$ which is obtained from the 
observed galaxy periodicity of $128\;{\rm Mpc}\;h^{-1}$. 

\medskip

In order to test our numerical code we have 
first restricted ourselves to the standard cosmology where known analytical 
solutions exist. The field equations have been solved by means of a 
fourth-order scheme with an adaptive stepsize control.
We have performed the integration of the equations with respect to 
time and also with respect to the variable $\alpha$. The time integration 
produces relative errors on the dynamical variables, like the scale factor, 
energy density, and Hubble parameter which are of the order of $10^{-8}$. 

The choice of the variable $\alpha$  as integration parameter allows us 
to explore the evolution at very early stages ($\alpha<0$ region) 
while keeping the relative errors small.
 Indeed we stop the integration at  
$\alpha\sim -25$, a bit beyond the value at which the nucleosynthesis 
takes place, the calculations having started at $\alpha=0$.

While the integration of equations for flat ($k=0$) and 
hyperbolic ($k=-1$) Universes can be performed for an arbitrarly large value 
of $\alpha$, for a closed Universe ($k=1$) the integration makes only sense 
for $\alpha \leq \alpha_{\rm max}$. The limiting value corresponds to the 
maximum size reached by the Universe and beyond which it 
starts recollapsing.  
We mention that the regions $\alpha > \alpha_{\rm max}$ are indeed not 
very interesting, first because the physics associated with them can be 
inferred from the $\alpha < \alpha_{\rm max}$ branch and then because 
no observational bounds arise from that region.

\vskip .3cm

We use the Hamiltonian constraint as a test on the accuracy of the 
procedure. It is applied at every 
integration step and implemented by defining the {\it deviation} parameter 
(see the Appendix) as,  
\be \label{lambda}
\lambda := \frac{ (H/H_0)^2 - \left( \Omega -1\right) e^{-2\alpha}}
{E/e_c}\,\,\,\,.
\ee
For an infinite accurate integration this parameter would
equal one and the deviation from this value indicates the degree 
to which Eq. (\ref{ham}) fails to be satisfied.

 Figure \ref{f:fig1_gal} shows the correlation between the relative 
errors found for the cosmological variables and that for the Hamiltonian constraint (i.e., in 
$\lambda$) as a function of time for the Einstein--de Sitter Universe. 
The deepest peaks (infinite precision) 
indicate the location of initial data. 
Only for convenience, in such points the precision has been arbitrarily 
set to be $\sim 10^{16}$.

 We have also verified this type of correlation in closed and 
hyperbolic Universes. Because, no analytical solutions are known for the oscillating 
models, we use Eq. (\ref{lambda}) 
systematically (the ``internal test'') in order to verify the accuracy of the 
results. 

\bigskip
\section{Numerical results}
\bigskip

\indent As we will see below, for any given value of $\Omega$ it is possible 
to fix $\Omega_{\rm bar}$ such that the correct value of the $^4$He abundance 
is recovered. 
Figure \ref{f:fig2_gal} shows the behavior of $\phi$ in some 
range of past and future epochs with $\Omega=1$ and the corresponding
 $\Omega_{\rm bar} \sim 0.021$.
 As the scale factor 
increases, the amplitude is damped due to redshifts while the 
frequency of oscillation (with respect to $\alpha$), which is inversely 
proportional to $H^2$ \cite{H^2}, grows (see Fig. \ref{f:fig3_gal} for the 
behavior of the Hubble parameter). 
The amplitude and oscillation frequency will reach maximum and 
minimum values respectively as the 
scale factor decreases before entering into a stage 
(discussed below) in which the scalar field almost vanishes.
 This includes a late era in which matter is 
dominant over the scalar field energy density and an earlier one 
in which this is the opposite (see Fig. \ref{f:fig4_gal}).

 We introduce effective energy densities by 
\bea
E_\xi &:=& 12\xi \phi P_\phi P_\alpha \frac{G_{\rm eff}}{G_0}\ ,\\
E_\phi &:=& \frac{G_{\rm eff}}{G_0}\left[\frac{1}{2}{P_\phi}^2 + 
m^2\phi^2 \right] \ ,\\
E_{\rm mat} &:=& \frac{G_{\rm eff}}{G_0} e \ .
\eea

\noindent Figure \ref{f:fig4_gal} 
shows how these fractions of the total effective energy density 
(depicted in Fig. \ref{f:fig5_gal}) vary with $\alpha$. 

We emphasize that $E_\xi$ (a ``coupling energy density'') is not positive 
definite and since it 
contributes to the total effective energy density $E$ [see Eq. (\ref{ene1})], 
there are some regions where the 
fraction $E_\xi/E$ is negative (dash-dotted line) while 
 $E_\phi/E$ and $E_{\rm mat} /E$ exceed one 
(dashed and solid lines, respectively). In particular, when comparing 
Figs. \ref{f:fig2_gal} and \ref{f:fig4_gal} we note that 
when the amplitude $|\phi|$ is large (in the region 
$\alpha\in [-4,-3]$), the net scalar-field's contribution to 
$E$ represented by 
$E_\xi + E_\phi$ is small because the negative ``energy density'' 
$E_\xi$ compensates the contribution from $E_\phi$. Indeed, it is 
the matter contribution $E_{\rm mat}$ which becomes dominant in this 
era.

When the scale factor is still small ($\alpha\in [-8,-5]$), 
$|\phi|$ falls down dramatically by 
entering what we called the ``fine-tuning era'' (see Sec. IV A) 
where $E_{\rm mat}$ dominates completely. 
The effective gravitational constant which was reduced in $\sim 38\% $ 
of its current value $G_0$ during the maxima of $|\phi|$, recovers its normal
value 
again (Fig.\ref{f:fig6_gal}). Finally, 
 for still smaller $\alpha$, the scalar field becomes dominant again, 
resulting in $G_{\rm eff}\rightarrow 0$ as we approach the Big Bang 
singularity. The Hubble parameter decreases monotonically in the early era 
($\alpha\in (-\infty,3]$) (Fig. \ref{f:fig3_gal}).

\subsection{The nucleosynthesis bound and ``fine-tuning''}
\bigskip

\indent In previous investigations, Crittenden and Steinhardt \cite{CritStein} 
have brought attention to a couple of much more severe constraints on the 
oscillating models 
that those considered before by Morikawa \cite{Morik} and Hill {\it et al.} 
\cite{Hill}. The first of these constraints regarded the bound imposed by the 
Brans-Dicke parameter 
tests which are even more stringent that the one coming 
from the Viking radar echo experiments. However, the authors argue that 
it can be eluded by assuming a ``fortunate phase'' 
for the oscillation of the scalar field when the nonminimally 
coupling function varies quadratically with the changes of $\phi$. 
The second new bound they mentioned arises from the fact that prior to the onset 
of its oscillatory behavior, the amplitude of $\phi$ had to be 
small enough in order to prevent 
$|\Delta G|/G>0.4$ at nucleosynthesis as described by 
Accetta {\it et al.} \cite{Acetal}, but at the same 
time, the evolution of $\phi$ has to 
be such that its present amplitude is large enough to accomodate the 
value of ${\cal A}_0$ needed to generate the observed peaks in the deep pencil 
survey\cite{BEKS}. Crittenden and Steinhardt suggested that both conditions might 
be possible by ``fine-tuning'' the amplitude
 of $\phi$ at nucleosynthesis. This 
amplitude might then be amplified by the effect of the curvature during the 
matter-dominated era and afterwards damped by the redshift's effects. 

Although the nucleosynthesis bound can be thought as imposing 
a limit on the variation of $G$, this is not enterely precise. 
The true bound that comes from nucleosynthesis is a limit on the 
deviation of the expansion rate of the Universe from its value given 
by the standard cosmological model. There exists a very 
narrow region for which the expansion rate of the Universe and the 
transition rate for the weak interaction, which convert   
neutrons to protons, traduce themselves into a freezeout temperature 
that reproduces the 
observed $^4{\rm He}$ abundance (see \cite{Turner} for a review). 
In the standard 
cosmology framework, for a radiation dominated era, the ratio between 
those rates, is 
\be
\frac{\Gamma}{H} \sim 
\left(\frac{ \kappa T}{0.7 {\rm \,\,MeV}}\right)^3 \,\,\,\,.
\ee
Thus  the temperature at which those weak interactions freeze-out 
the $n/p$ ratio is 
$\kappa T_F \sim 0.7{\rm\,\, MeV}$. At this 
temperature, the neutron-proton-ratio is given by its equilibrium value
\be\label{npr}
\left(\frac{n}{p}\right) = {\rm exp}\left(-Q/T_F\right) \approx 1/6 \ ,
\ee
where $Q$ is the mass difference of neutrons and protons.
This ratio can decrease to $1/7$ if we take into account the ``natural''
neutron decay due to weak interactions. This value predicts approximately 
one $^4{\rm He}$ nuclei for each $16$H nuclei. Therefore 
the primordial abundance of $^4{\rm He}$-${\rm H}$ ratio 
gives $Y_P \approx 25\%$, i.e., very close to the observed 
value, varying in some small amount depending on the number of neutrino 
species considered.

Let us emphasize that one of the greatest achievements of the standard 
cosmology is the predicted light element abundance that we observe today. 
Therefore, the oscillating model would be considered as viable if it preserves 
also these nice features.

Because of the exponential relation (\ref{npr}), a small variation of the 
freezeout temperature will produce a large deviation of the observed 
neutron-proton ratio. Deviations of the freezeout temperature will arise 
as a result of a small change of the expansion rate. The greater the 
expansion rate, the greater the freezeout temperature and so larger 
the $^4{\rm He}$ abundance.
This is the main reason why alternative cosmological models 
can fail.

We showed in the previous section that the initial condition $\phi_0$ and the 
value of $\xi$ depends upon ${\cal A}_0$ , $\omega$, $\Omega$, and $\Omega_{\rm bar}$.
The parameters $\omega$ and ${\cal A}_0$ are fixed in order to satisfy the 
observed galaxy periodicity and the galaxy peak amplitude whereas
$\dot \phi_0=0$ prevents conflicts with the Viking experiments constraint. So  
we can try to adjust $\Omega$ or $\Omega_{\rm bar}$ in order to 
ensure the satisfaction of the nucleosynthesis bound. Initially we 
considered  fixing  $\Omega_{\rm bar}= 0.1$ and then adjusting $\Omega$. 
Another perhaps more natural possibility (see discussion) is to take a 
value of $\Omega$ (in particular $\Omega= 1$ as suggested by inflation) 
and then adjusting $\Omega_{\rm bar}$. As we will see below, we have done
this obtaining $\Omega_{\rm bar} \sim 0.021$.

In the process of integrating towards small values of $\alpha$ we 
find initially the following behavior: for some choices of $\Omega_{\rm bar}$ , 
the value of $\phi$ goes monotonically to $+\infty$, or to 
$-\infty$ for some others. This suggested to us that there exists some value of 
$\Omega_{\rm bar}$ at which the transition from one behavior to the other 
takes place. We found that for $\Omega =1$ this ``fine-tuned'' value turns
out to be $\Omega_{\rm bar} \sim 0.021$.
Figure \ref{f:fig7_gal} shows the behavior of $\phi$  
for three values of $\Omega_{\rm bar}$ about this point.

This transition point, traduces 
in a very special ``initial'' conditions for which the $\phi$ amplitude 
is ``squeezed'' to zero, before the onset of the oscillatory behavior. 
The search of that transition value is what we call 
{\it ``fine-tuning.''} Our numerical experience indicates that 
it is possible to generate  a finite region during which $\phi\sim 0$ 
before growing or decreasing
 monotonically (in the direction of decreasing $\alpha$) 
and that the extent of that region depends on the improvement of the 
``fine-tuning''.

Different ``experiments'' in ``fine-tunings''
 are depicted in Fig. \ref{f:fig8_gal}. 
The larger the plateau for which $G_{\rm eff} \rightarrow G_0$ 
(i.e., for which $\phi \rightarrow 0$) the closer 
the freezeout temperature is to the standard 
cosmology prediction $0.7\,\,{\rm MeV}$. 
We have found that 
$\Omega_{\rm bar}= 0.021 + \epsilon$ results in a freezeout temperature which agrees
with  $0.7\,\,{\rm MeV}$ and therefore giving a  
$^4{\rm He}$ abundance that approximates best the observed value (see Fig.
\ref{f:fig9_gal}).

In light of the 
extreme sensitivity of $T_F$ to the value of $\Omega_{\rm bar}$ we are not able 
to improve the ``fine-tuning" due to the limitations in the numerical precision. 
%This improvement would result in a plateau of small $\phi$-amplitude extended 
%to nucleosynthesis. However, even with a not infinte precise
However, even with a noninfinitely precise ``fine-tuning" the model 
is able to recover the $^4{\rm He}$ abundance from nucleosynthesis. 
Figure \ref{f:fig9_gal} shows the freezeout temperature 
($\sim 0.7\,\,{\rm MeV}$) predicted by the best ``fine-tuning" we explored.

The $\Omega = 1$  scenario (with $\Omega_{\rm bar}\sim 0.021$) corresponds to 
an age of the Universe of $\sim 0.8 H_0^{-1}$ (see Fig.\ref{f:fig10_gal}). 
 Figure \ref{f:fig11_gal} shows a typical curve of the Hubble 
parameter as a function of redshifts. 
The transition between the matter and radiation epochs is clearly 
appreciated in Fig. \ref{f:fig12_gal} [see also Eq. (\ref{inte})].

Only for completness we mention that, for example, if we choose 
$\Omega_{\rm bar}= 0.1$ and fine-tune $\Omega$ this results in 
$\Omega\sim 1.67$ which corresponds to 
a closed Universe. %with a $\Omega$ exceeding $\sim 67\%$ the critical value. 
Figure \ref{f:fig13_gal} shows that 
the age of the Universe based in this scenario is about $0.7 H_0^{-1}$. 
Somewhat younger than the age predicted by a Universe that includes only 
baryonic matter and radiation energy densities (a 
hyperbolic-standard-cosmology model) which corresponds to $\sim 0.9 H_0^{-1}$. 
%The maximal size reached by the Universe before starting its contraction 
%would be $\sim 2.5$ times its current size in a time of $\sim 4H_0^{-1}$ 
%from today. The Big Crunch 
%is then predicted at $\sim 5H_0^{-1}$ after the recollapse has started.
Figure \ref{f:fig14_gal} shows the freezeout temperature 
($\sim 0.8\,\,{\rm MeV}$) predicted by the best ``fine-tuning" we explored
in the $\Omega_{\rm bar} =0.1$, \,$\Omega \sim 1.67$  scenario. 
This temperature is very close to the one predicted by the 
standard cosmology ($\sim 0.7\,\,{\rm MeV}$). This examplifies the fact that 
for any choice of $\Omega$ (or $\Omega_{\rm bar}$) the freezeout temperature
of standard cosmology can be recovered by adjusting the other parameter 
$\Omega_{\rm bar}$ (or $\Omega$).  

\bigskip
\section{Conclusion and Discussion}
\bigskip

At first sight the model could not be less appealing: a ``fortunate phase''
and an incredibly precise ``fine-tuning."
However, we must judge it in the proper context by contemplating the alternatives.
\smallskip

 First the observed periodicity might be a statistical fluke, then of course that would
be the end of the story. However as we argued in the Introduction it would be quite
a coincidence that we just happened to
look at the couple of directions that exhibit such patterns. In any event this is a
matter that only further observations will answer.

\smallskip

 Next we have the ``spontaneous breaking of the cosmological principle'' \cite{Bun}, clearly
 a major departure form our cosmological conceptions which besides that, requires
something even more fortunate than the ``fortunate phase'' of the oscillating $G$
models: We happened to be ``fortunate'' enough to be born in a galaxy
which happens to lay in the middle of a concentric collection of shells of
maximal galaxy density.
\smallskip

 And finally the {\it 
galactic luminosity oscillation} model, which as we said at the beginning,
more than a model is a vaguely specified scenario
which nevertheless seems to require at least two new hypotheses: a new
type of star cooling mechanism, and also  (as the other alternatives do) a driving 
oscillating cosmological scalar field to turn on and off that mechanism
periodically in cosmic time. Needless to say that once
the scenario is implemented with a specific  model, unforsen
new bounds might also have to be overcome.

\smallskip

In this light, the oscillating $G$ model does not look as clearly dismissable.
Furthermore, while it is true that the fortunate phase will
have to remain such, for the other problem, ``the fine-tuning,'' we will argue 
below that there are scenarios in which this problem is not present.

To start, we must stress that while the ``fine-tuning" is completely
unnatural when approached, as we have, from the present to the past, 
when looked from the opposite and more natural direction, the situation is
quite different. In fact all that seems to be required is for some
mechanism
to drive the scalar field to an extremely low value before the era of nucleosynthesis,
for then our calculations show the field will remain at that value up to and beyond
that  era so will have $G_{\rm eff} \approx G_0$
and then the success of ``Big Bang nucleosynthesis'' will be recovered naturally.
The amplitude $|\phi|$ will later be amplified by the curvature coupling
precisely before the onset of the oscillatory behavior.
\smallskip

Thus it is possible that starting from an arbitrary value of  $\phi$ near the 
Big Bang, a mechanism (no different from that required to solve the other problems of
the standard Big Bang model, including inflation itself) would
drive the scalar field to a value near zero, at which it will remain  
until just before 
$H\approx m$ when the amplification and then oscillations would occur.
This type of scenario might be combined with the inflationary prediction 
$\Omega=1$ and a corresponding ``fine-tuning" on $\Omega_{\rm bar}$. This has been done 
and the ``fine-tuning" yielded a value $\Omega_{\rm bar}\sim 0.021$, surprisingly in the 
very narrow range $0.016 \leq \Omega_{\rm bar} \leq 0.026$ that results in a successful 
nucleosynthesis of the light elements other than $^4 {\rm He}$.
 But the point is that  this would
really be no ``fine-tuning" at all (if looked in the right perspective), 
because
all that it will mean is that, given the physical constant {\it precise
values}, the inflationary mechanism will ensure that
 the energy densities of the 
various components, scalar field and ordinary matter, add up to 
$\Omega =1$, which 
will correspond then to a Universe at our time with {\it precise} values of the
densities, expansion rate, etc. In particular the precise current value of the 
scalar field amplitude and phase arises from a particular precise value of the 
parameters at early times, among them the baryon content of the Universe. 
The fact 
that the corresponding baryonic density today turns out to lay in a very 
narrow range consistent with the light element's nucleosynthesis 
suggests that the model should be taken seriously. 
In this respect we would like to point out that the {\it a priori}
 probability for this 
happening just by chance is about 1 in 100 (the range $[0.016, 0.026]$ which is
observationally allowed for $\Omega_{\rm bar}$ represents 
approximately 1 part in 100 inside the numerical range allowed in 
 principle $[0, 1]$).

In view of the previous arguments we may conclude by 
saying that the ``fine-tuning" should be seen as simply recovering 
the ``initial''  conditions corresponding to the observational data and that,
moreover, this has 
 resulted in the specific prediction $\Omega_{\rm bar}\sim 0.021$. 
Therefore, 
this is certainly the most atractive of all the models considered in order 
to explain the observed periodicity in the galactic distribution, and
should also be considered as a missing mass model with the scalar field
playing the role of dark matter which is, however, indirectly observable
in the oscillation of the galactic distribution. 

%The latter procedure is therefore much more advantageous that the time 
%integration and it provides the 
%possibility of exploring numerically regions of a very small scale factor that%could be interesting in another context like inflation \cite{SteinWill}.
\begin{appendix}
\section*{Dimensionless form of field equations}
\bigskip
\subsection{Einstein's dynamical equation}
\bigskip

It is easy to check that when restoring factors of $c$ and introducing 
characteristic lengths, Eq. (\ref{dyn1}) becomes

\be \label{DynFsd} 
\dot{\tilde P_\alpha}  = {\tilde P_\alpha}^2 + 
\frac{1}{2}\left[ \tilde E +  \tilde S \right] 
\,\,\,\,,
\ee

with
\bea \label{dimtim}
\tilde t &:=& tH_0 \,\,\,\,,\\
\label{dimpal}
\tilde P_\alpha &:=& -\dot{\tilde \alpha}  \,\,\,\,,\\
\tilde E &:=& \frac{E}{e_c} \,\,\,\,\,,\\
\tilde S &:=& \frac{S}{e_c} \,\,\,\,\,,\\
\label{dencrit}
e_c &:=& \frac{3c^2 H_0^2}{8\pi G_0}\,\,\,\,.
\label{dimPalpha}
\eea
Here a dot over a tilde means derivation with respect the dimensionless time 
$\tilde t$.

\bigskip
\subsection{Scalar field's equation of motion}

Using (\ref{dimtim})--(\ref{dimPalpha}) and introducing
\be \label{dimfi}
\tilde P_\phi := \dot{\tilde \phi} \,\,\,\,
\ee
Eq. (\ref{seq2}) can be written
\
\be \label{DynSFFsd} 
\dot{\tilde P_\phi}  
=  3 \tilde P_\alpha \tilde P_\phi - \frac{3}{8\pi} {\tilde V}'(\phi) - 
6\xi \,\phi \left[ \tilde S - \tilde E \right] \,\,\,\,.
\ee
The scalar field $\phi$ now turns to be a dimensionless 
quantity and
\be
\tilde V(\phi) := \frac{V(\phi)}{e_c} \,\,\,\,\,.
\ee
Thus, in the case of a scalar potential $V(\phi)= \Lambda \phi^n$, 
$\Lambda$ has units of 
energy density. In particular if $\Lambda:= m^2$ and $n=2$, the 
dimensionless harmonic frequency of oscillation is given by
\be
\tilde \omega := \frac{\omega}{H_0}= \sqrt{\frac{3}{4\pi}} \tilde m \,\,\,\,.
\ee
Moreover, with such a choice of units, the 
coupling constant $\xi$ is also dimensionless. 
Finally, the dimensionles source terms read explicitly 
\bea
\label{Ssd}
\tilde S &=& \frac{3\tilde G}{1+ 192\pi\tilde G \xi^2\,\phi^2}
\left[  \tilde p + \frac{4\pi}{3} 
{\tilde{P_ \phi}}^2 - \tilde 
V(\phi) \right. \nonumber \\
 & +& \left. \frac{32\pi}{3}\xi \left(\phi {\tilde P_\phi} {\tilde P_\alpha} + 
{\tilde P_\phi}^2 -\frac{3}{8\pi}\phi\, {\tilde V}'(\phi)\right) \right] 
\nonumber \\
\label{SFTFsd}
 &+& \frac{192\pi \tilde G\,\xi^2 \phi^2 \,\tilde E}
{1+ 192\pi \tilde G \xi^2 \,\phi^2} \,\,\,\,,\\
\label{EtotFsd}
\tilde E &=& \tilde G\, \left[\tilde e + \frac{4\pi}{3} {\tilde {P_\phi}}^2 + 
\tilde V(\phi) 
+32\pi \xi \phi \tilde{P_\phi} {\tilde P_\alpha} \right]  \,\,\,\,,\\
\label{Gtilsd}
\tilde G &=& \left[ 1+ 16\pi\,\xi\,\phi^2\right]^{-1}
\,\,\,\,.
\eea

%It's to be noticed that equation (\ref{DynSFFsd}) with source terms 
%(\ref{Ssd}), (\ref{EtotFsd}) and (\ref{Gtilsd}) can be recognized as the 
%dimensionless form of equation (2.4b) of Steinhardt \& Will \cite{SteinWill} 
%with the choice $f= (1+ 16\pi\xi\phi^2)/16\pi$, being $f$ their non-minimal 
%coupling function.

Equations (\ref{DynFsd}), (\ref{dimpal}), (\ref{dimfi}), (\ref{DynSFFsd})  
with source terms (\ref{Ssd})--(\ref{Gtilsd}) and the equation of 
conservation of energy (\ref{ConsEFsd}) 
are then the complete set of equations best suited to be solved numerically.
With a trivial manipulation we write these equations in terms of 
derivatives with respect to $\alpha$ instead of $\tilde t$.

\subsection{The Hamiltonian constraint and the deviation parameter}
\bigskip

\indent With these notations, the dimensionless form of the 
Hamiltonian constraint (\ref{ham}) reads
\be \label{HamFsd}
{\tilde P_\alpha}^2 + \frac{k}{a^2 H_0^2}= \tilde E  \,\,\,\,.
\ee 
Now, at $t=t_0$ this becomes
\be \label{constinit}
1 + \frac{k}{a^2_0 H_0^2}= \tilde E_0 \,\,\,\,.
\ee
We can now replace $k$ in Eq. (\ref{HamFsd}) in terms of hereabove 
initial conditions to get
 
\be \label{HamFsd3} 
{\tilde P_\alpha}^2 + \left( \tilde E_0 -1\right) e^{-2\alpha} - \tilde E= 
0 \,\,\,\,,
\ee 
where we have employed our variable $\alpha$ instead of $a$.

We can introduce also the dimensionless 
deceleration parameter $\tilde q(t)$ in terms of sources
\be \label{constinit2}
\tilde q(t)= \frac{1}{2 {\tilde P_\alpha}^2}  \left[ \tilde E +  \tilde S\right] \,\,\,\,\,.
\ee
At $t=t_0$, we can also rewrite (\ref{constinit2}) in the form
\be \label{iceq2} 
\tilde E_0 = 2\tilde q_0 - \tilde S_0 \,\,\,\,\,\,\,.
\ee

\noindent Moreover, the deviation parameter introduced in (\ref{lambda}) take 
the form
\be \label{lambda2}
\lambda := \frac{ {\tilde P_\alpha}^2 - \left( \Omega -1\right) e^{-2\alpha}}
{\tilde E} \,\,\,\,,
\ee
where we stress ${\tilde P}_\alpha \equiv -H(t)/H_0$ and 
$\Omega \equiv \tilde E_0$. 

\bigskip
\subsection{Conservation equations of matter and radiating fields}
\bigskip

Equation (\ref{claw}) becomes

\be \label{ConsEFsd} 
\dot{\tilde e}_i = 3 (\tilde e_i+ \tilde p_i) \tilde P_\alpha \,\,\,\,.
\ee

\end{appendix}

%\vskip 4cm
%\centerline{\Large \bf Captions of figures}
\newpage

\begin{figure*}
%\picplace{8cm}
%\vspace{1cm}
%\psfig{figure=../../Numeric.dir/fig1_gal.ps,angle=-90,height=14cm,width=5.5in}
%\hspace*{-2.5in}
\caption[]{\label{f:fig1_gal}
%small
 Relative errors as a function of 
dimensionless time: solid lines(superposed): 
${\rm log}_{10}|\left( a(t)_{\rm an}-a(t)_{\rm num}\right) /a(t)_{\rm an}|$, 
\,\,${\rm log}_{10}|\left( \dot a(t)_{\rm an}-\dot a(t)_{\rm num}\right) 
/\dot a(t)_{\rm an} |$;
 dashed line: 
${\rm log}_{10}|\left( \alpha(t)_{\rm an}-\alpha(t)_{\rm num}\right) /
\alpha(t)_{\rm an} |$; dash-dotted line:
${\rm log}_{10}|\left(P_\alpha(t)^{\rm an}-P_\alpha(t)^{\rm num}\right) /
P_\alpha^{\rm an} |$; 
dotted line: ${\rm log}_{10}|\left( (ea^3)_{\rm an}- (ea^3)_{\rm num} \right) /
(ea^3)_{\rm an} |$;
dot-dashed line (bottom): ${\rm log}_{10}|1- \lambda|$. The initial 
conditions located at $t_0:=0.5H_0^{-1}$ .}
\end{figure*}

\begin{figure*}
%\picplace{2cm}
%\vspace{1cm}
%\psfig{figure=../../Numeric.dir/fig2_gal.ps,angle=-90,height=14cm,width=5.5in}
%\hspace*{-2.5in}
\caption[]{\label{f:fig2_gal}
%small
Fine-tuned scalar field amplitude as a function of ${\rm ln}[a/a_0]$ for 
a flat Universe ($\Omega=1$) with
$\Omega_{\rm bar}= 0.021\,\,012\,\,641\,\,182\,\,345$ and ${\cal A}_0= 0.5$ 
at the onset of oscillations. At present time ($\alpha= 0$) the initial 
amplitude is $\phi_0\sim 3.288\times 10^{-3}$ and $\xi\sim 6.267$. Computations were stopped at $\alpha\sim 1.5$ .}
\end{figure*}

\begin{figure*}
%\picplace{8cm}
%\vspace{1cm}
%\psfig{figure=../../Numeric.dir/fig3_gal.ps,angle=-90,height=14cm,width=5.5in}
\caption[]{\label{f:fig3_gal}
%small
The Hubble parameter obtained by using the same initial values 
as in Fig. \ref{f:fig2_gal} and $H(\alpha=0)= H_0$.}
\end{figure*}

\begin{figure*}
%\picplace{8cm}
%\vspace{1cm}
%\psfig{figure=../../Numeric.dir/fig4_gal.ps,angle=-90,height=14cm,width=5.5in}
%\hspace*{-2.5in}
\caption[]{\label{f:fig4_gal}
%small
Starting from the same initial values as in Fig. \ref{f:fig2_gal}, 
this figure depicts the fractions of effective 
energy densities: $E_{\rm mat}$ 
(solid line), $E_\phi$ (dashed line) and, 
$E_\xi$ (dash-dotted line) (see text for definitions).}
\end{figure*}

\begin{figure*}
%\picplace{8cm}
%\vspace{1cm}
%\psfig{figure=../../Numeric.dir/fig5_gal.ps,angle=-90,height=14cm,width=5.5in}
%\hspace*{-2.5in}
\caption[]{\label{f:fig5_gal}
%small
The evolution of the total effective energy density resulting from the 
numerical integration of the field equations of Sec. II (see also the Appendix), 
with initial values as in Fig. \ref{f:fig2_gal}.}
\end{figure*}

\begin{figure*}
%\picplace{8cm}
\vspace{1cm}
%\psfig{figure=../../Numeric.dir/fig6_gal.ps,angle=-90,height=14cm,width=5.5in}
%\hspace*{-2.5in}
\caption[]{\label{f:fig6_gal}
%small
Behavior of the effective gravitational ``constant'' 
in units of Newton's constant $G_0$. The initial 
values have been chosen as in Fig.\ref{f:fig2_gal}.}
\end{figure*}

\begin{figure*}
%\picplace{8cm}
%\vspace{1cm}
%\psfig{figure=../../Numeric.dir/fig7_gal.ps,angle=-90,height=14cm,width=5.5in}
%\hspace*{-2.5in}
\caption[]{\label{f:fig7_gal}
%small
Behavior of the scalar field for three different values of 
$\Omega_{\rm bar}$ within the $\Omega=1$ scenario. 
The solid line represents the best ``fine-tuning"
obtained with $\Omega_{\rm bar}= 0.021\,\,012\,\,641\,\,182\,\,345$ (see Fig.\ref{f:fig2_gal});
the dashed line for $\Omega_{\rm bar}= 0.022$, and the dash-dotted line with 
$\Omega_{\rm bar}= 0.020$.}
\end{figure*}

\begin{figure*}
%\picplace{8cm}
%\vspace{1cm}
%\psfig{figure=../../Numeric.dir/fig8_gal.ps,angle=-90,height=14cm,width=5.5in}
%\hspace*{-2.5in}
\caption[]{\label{f:fig8_gal}
%small
Effective gravitational ``constant'' 
in units of Newton's constant $G_0$ for different fine-tunings: solid line 
obtained with $\Omega_{\rm bar}= 0.021\,\,012\,\,641\,\,182\,\,345$,
dashed line for $\Omega_{\rm bar}= 0.021\,\,012\,\,641\,\,18$ dash-dotted line with 
$\Omega_{\rm bar}= 0.021\,\,012\,\,6$.}
\end{figure*}

\begin{figure*}
%\picplace{8cm}
%\vspace{1cm}
%\psfig{figure=../../Numeric.dir/fig9_gal.ps,angle=-90,height=14cm,width=5.5in}
%\hspace*{-2.5in}
\caption[]{\label{f:fig9_gal}
%small
Expansion (solid line) and transition (dashed line) rates in terms of 
the blackbody temperature. The asterisk depicts the freezeout temperature 
$\sim 0.7\,\, {\rm MeV}$ at which nucleosynthesis takes place as predicted 
by the standard cosmological models. The cross-point of 
both curves indicates the corresponding freezeout temperature 
$\sim 0.7\,\,{\rm MeV}$ for the oscillating model of previous figures.}
\end{figure*}

\begin{figure*}
%\picplace{8cm}
%\vspace{1cm}
%\psfig{figure=../../Numeric.dir/fig10_gal.ps,angle=-90,height=14cm,width=5.5in}
%\hspace*{-2.5in}
\caption[]{\label{f:fig10_gal}
%small
Scale factor of the oscillating model 
of previous figures in units of its value today as a function of 
cosmic time (in units of $H_0^{-1}$). The present time $t_0$ has been placed 
at $t= 1H_0^{-1}$.}
\end{figure*}

\begin{figure*}
%\picplace{8cm}
%\vspace{1cm}
%\psfig{figure=../../Numeric.dir/fig11_gal.ps,angle=-90,height=14cm,width=5.5in}
%\hspace*{-2.5in}
\caption[]{\label{f:fig11_gal}
%small
The Hubble parameter as a function of redshifts. The initial 
conditions of Figs. \ref{f:fig3_gal} and \ref{f:fig10_gal} determine the 
initial value $H/H_0= 1$ at $z(t_0)= 0$. }
\end{figure*}

\begin{figure*}
%\picplace{8cm}
%\vspace{1cm}
%\psfig{figure=../../Numeric.dir/fig12_gal.ps,angle=-90,height=14cm,width=5.5in}
%\hspace*{-2.5in}
\caption[]{\label{f:fig12_gal}
%small
The combined matter-radiation (solid line) 
and pure radiation (dashed line) energy densities 
with initial values as in Fig. \ref{f:fig2_gal}.}
\end{figure*}

\begin{figure*}
%\picplace{8cm}
%\vspace{1cm}
%\psfig{figure=../../Numeric.dir/fig13_gal.ps,angle=-90,height=14cm,width=5.5in}
%\hspace*{-2.5in}
\caption[]{\label{f:fig13_gal}
%small
Scale factor for the closed oscillating Universe with $\Omega\sim 1.67$ 
and $\Omega_{\rm bar}= 0.1$ (solid line). The dashed line corresponds to the 
case with no scalar 
field (standard cosmology). 
As in Fig.\ref{f:fig10_gal}, the present time $t_0$ has been placed at 
$t= 1H_0^{-1}$.}
\end{figure*}

\begin{figure*}
%\picplace{8cm}
%\vspace{1cm}
%\psfig{figure=../../Numeric.dir/fig14_gal.ps,angle=-90,height=14cm,width=5.5in}
%\hspace*{-2.5in}
\caption[]{\label{f:fig14_gal}
%small
Similar to Fig.\ref{f:fig9_gal} for the closed Universe with 
$\Omega_{\rm bar}= 0.1$ and $\Omega\sim 1.67$. 
Here the predicted freezeout temperature closely agrees 
with $\sim 0.7\,\, {\rm MeV}$.}
\end{figure*}


\begin{thebibliography}{}
\bibitem{H^2}This frequency can be obtained directly from 
equation (\ref{seq2}) when changing the integration 
variable from $t$ to $\alpha$.
\end{thebibliography}
\end{document}